\documentclass[useAMS,usenatbib]{mn2e}

\usepackage{epsfig}
\usepackage{graphicx}
\usepackage[latin1]{inputenc}
\def\simlt{\ \raise -2.truept\hbox{\rlap{\hbox{$\sim$}}\raise5.truept   %
\hbox{$<$}\ }}
\def\simgt{\ \raise -2.truept\hbox{\rlap{\hbox{$\sim$}}\raise5.truept   %
\hbox{$>$}\ }}                                                          %

\def\be{\begin{equation}}
\def\ee{\end{equation}}
\def\newline{\hfil\break}

\def\la{\mathrel{\hbox{\rlap{\hbox{\lower4pt\hbox{$\sim$}}}\hbox{$<$}}}}
\def\ga{\mathrel{\hbox{\rlap{\hbox{\lower4pt\hbox{$\sim$}}}\hbox{$>$}}}}

\title[Non-thermal SZE]{Effect of the non-thermal Sunyaev-Zel'dovich Effect on the temperature determination of galaxy clusters}

\author[P. Marchegiani and S. Colafrancesco]
{P. Marchegiani$^{1}$\thanks{E-mail: Paolo.Marchegiani@wits.ac.za} and S. Colafrancesco$^{1}$\thanks{E-mail: Sergio.Colafrancesco@wits.ac.za}\\
$^{1}$School of Physics, University of the Witwatersrand, Private Bag 3, 2050-Johannesburg, South Africa
}
\begin{document}

\date{Accepted 2017 May 18. Received 2017 April 25; in original form 2017 February 13}

\pagerange{\pageref{firstpage}--\pageref{lastpage}} \pubyear{2017}

\maketitle

\label{firstpage}

\begin{abstract}
A recent stacking analysis of Planck HFI data of galaxy clusters (Hurier 2016) allowed to derive the cluster temperatures by using the relativistic corrections to the Sunyaev-Zel'dovich effect (SZE). However, the temperatures of high-temperature clusters, as derived from this analysis, resulted to be basically higher than the temperatures derived from X-ray measurements, at a moderate statistical significance of $1.5\sigma$. This discrepancy has been attributed by Hurier (2016) to calibration issues. In this paper we discuss an alternative explanation for this discrepancy in terms of a non-thermal SZE astrophysical component. We find that this explanation can work if non-thermal electrons in galaxy clusters have a low value of their minimum momentum ($p_1\sim0.5-1$), and if their pressure is of the order of $20-30\%$ of the thermal gas pressure. Both these conditions are hard to obtain if the non-thermal electrons are mixed with the hot gas in the intra cluster medium, but can be possibly obtained if the non-thermal electrons are mainly confined in bubbles with high content of non-thermal plasma and low content of thermal plasma, or in giant radio lobes/relics located in the outskirts of clusters. In order to derive more precise results on the properties of non-thermal electrons in clusters, and in view of more solid detections of a discrepancy between X-rays and SZE derived clusters temperatures that cannot be explained in other ways, it would be necessary to reproduce the full analysis done by Hurier (2016) by adding systematically the non-thermal component of the SZE.
\end{abstract}


\section{Introduction}

The Sunyaev-Zel'dovich effect (SZE) is the distortion of the spectrum of the Cosmic Microwave Background (CMB) photons produced by Inverse Compton Scattering off the electrons in the hot plasma in galaxy clusters (Zel'dovich \& Sunyaev 1969). It is well known that it can be used as a powerful probe of the thermal plasma properties in galaxy clusters (see, e.g., Pointecouteau et al. 1998, Birkinshaw 1999), and also of the non-thermal plasma, like relativistic electrons populations (e.g. Colafrancesco et al. 2003, Colafrancesco et al. 2011).

Recently, Hurier (2016; hereafter H16) has analyzed the Planck HFI data of galaxy clusters (Planck Collaboration 2011) to derive the temperature of a large sample of galaxy clusters using the relativistic corrections of the thermal SZE that, being dependent on the cluster temperature, can break the degeneracy between the cluster temperature and the optical depth that is present in the non-relativistic expression of the SZE (see Colafrancesco et al. 2003 for a complete discussion). By stacking the Planck data of thousands of clusters, H16 found the signature of the relativistic corrections to the thermal SZE with the highest statistical significance obtained up to date. H16 has been able also to determine the average temperature of temperature-binned clusters. H16 found that for high temperature clusters the 545 GHz Planck channel shows a higher emission compared to low-temperature clusters, and that the temperature that can be estimated from the SZE is basically higher than the temperature estimated from X-ray observations, even though with a moderate statistical significance ($\simlt 1.5 \sigma$). Since the effects of other contaminations (like, e.g., kinematic SZE, background and foreground contaminations, radio discrete sources) have been considered and eliminated from the analysis, H16 attributed this discrepancy to the effect of calibration uncertainties.

However, another physical possibility we want to discuss in this paper is that this discrepancy can be due to the presence of a non-thermal component of the SZE, as produced by non-thermal relativistic electrons, like the electrons that produce the radio halos observed in many galaxy clusters especially with high temperatures (e.g. Feretti et al. 2012). 
It has been found (Colafrancesco et al. 2003, Colafrancesco et al. 2011) that the deviations from the spectral shape of a pure thermal SZE due to a non-thermal electron contribution are more evident at high frequencies; since the discrepancy between the SZE and the X-ray determined temperatures in H16 are observed mainly in the 545 GHz channel and for high temperature clusters, where the non-thermal phenomena are usually more relevant (e.g. Feretti et al. 2012), it is natural therefore to suppose that the non-thermal SZE can be the origin of this discrepancy. 
For example,  Colafrancesco et al. (2011) showed that in the Bullet cluster a single gas temperature fit to the SZE data provides a best-fit value of $k_B T_e\sim22$ keV, whereas considering the additional contribution of a non-thermal SZE produced by non-thermal electrons with properties similar to the ones producing the radio halo, it is possible to obtain a better fit with a thermal electron population with temperature of the order of the X-ray derived one, i.e. $k_B T_e\sim14$ keV, and with the the addition of a non-thermal electron population providing its main contribution to the SZE at high frequencies.

In this paper we explore, therefore, this possibility. To this aim, as a preparatory work in view of future and more significant detections of this discrepancy, we attribute the discrepancy suggested by the H16 results to the contribution of a non-thermal SZE, and we derive the properties of the electrons that are required to produce this effect. Finally, we discuss if these physical requirements are reasonable and we discuss their possible impact on the physics of galaxy clusters, as well as other possible explanations of this discrepancy.

\section{Methods}

Following the general approach of Colafrancesco et al. (2003), the SZE can be written in the general form
\begin{equation}
\Delta I (x) = I_0 y g(x)
\end{equation}
where $I_0=2(k_B T_0)^3/(hc)^2$, $x=(h\nu)/(k_BT_0)$, and the Compton parameter $y$, in its general form, is given by
\begin{equation}
y=\frac{\sigma_T}{m_e c^2} \int P_e d\ell \;,
\label{compt.par}
\end{equation}
where $P_e$ is the electrons pressure, that in the general case can be due to thermal or non-thermal electrons, and the integral is performed along the line of sight $\ell$. The function $g(x)$ contains the spectral dependence of the SZE, and depends on the electron temperature for the thermal SZE, and on the properties (spectral shape, minimum electron momentum) of non-thermal electrons for the non-thermal SZE.

We notice that H16 used an approach where the thermal SZE is given by the sum of the non-relativistic part, where the function $g(x)$ is independent on the cluster temperature, and of relativistic corrections, that instead depend on the cluster temperature (see, e.g., Nozawa et al. 2000). 
In the following, we use a different approach (Wright 1979, En\ss lin \& Kaiser 2000, Colafrancesco et al. 2003), where the SZE is derived following a full relativistic calculation that can be applied to both the thermal and the non-thermal SZE. The two approaches have been shown to be equivalent within the ranges of validity of the fitting formulae used for the relativistic corrections (see, e.g., Boehm \& Lavalle 2009). Therefore, in our approach the total SZE can be written as
\begin{equation}
\Delta I_{tot}(x)  =  \Delta I_{th}(x) + \Delta I_{nt}(x) \;,
\end{equation}
where the quantity $\Delta I_{th}$ includes both the non-relativistic part and the relativistic corrections to the thermal SZE, and $\Delta I_{nt}$ is the non-thermal SZE.

With these assumptions, it is possible to calculate the spectral shape of the function $g(x)$, that in the general form, and working at the first order in the optical depth $\tau$, can be written as:
\begin{equation}
g(x)=\frac{m_e c^2}{\langle k_B T_e \rangle}[j_1(x)-j_0(x)] \;.
\label{general.g}
\end{equation}
In this expression, the quantity $\langle k_B T_e \rangle$ is the analogous of the electron temperature for a general electron population (Colafrancesco et al. 2003). For thermal electrons the equivalence $\langle k_B T_e \rangle = k_B T_e$ holds, while for non-thermal electrons we find:
\begin{equation}
\langle k_B T_e \rangle = \frac{1}{3}\int_0^\infty dp f_e(p) p \beta m_e c^2 \;,
\end{equation}
where $f_e(p)$ is the normalized spectrum of the electrons written as a function of their normalized momentum $p=\beta\gamma$.

In eq.(\ref{general.g}), the quantity $j_0(x)$ is the spectrum of the CMB normalized to $I_0$,
\begin{equation}
j_0(x)=\frac{x^3}{e^x-1} \;,
\end{equation}
and the quantity $j_1(x)$ is given by
\begin{equation}
j_1(x)=\int_{-\infty}^{+\infty} j_{0}(xe^{-s}) P_1(s) ds \;,
\end{equation}
where the function $P_1(s)$ is given by
\begin{equation}
P_1(s)=\int_0^\infty f_e(p) P_s(s,p) dp \;,
\end{equation}
and where the function $P_s(s,p)$ includes the full relativistic physics of the inverse Compton scattering process (see, e.g., En\ss lin \& Kaiser 2000, Colafrancesco et al. 2003). 
In the following we use for the non-thermal electrons spectrum a single power law shape with a minimum momentum $p_1$:
\begin{equation}
f_e(p) \propto p^{-s_e} \qquad p\geq p_1 \;.
\end{equation}
It has been found (Colafrancesco et al. 2003) that in this case the function $g(x)$ depends on both the values of $s_e$ and $p_1$.

Following these expressions, we can write the total SZE as:
\begin{equation}
\frac{\Delta I_{tot}(x)}{I_0} = g_{th}(x;T_e)y_{th} + g_{nt}(x;s_e,p_1)y_{nt} \;,
\end{equation}
where we define $g_{th}$ and $g_{nt}$ as the function $g(x)$ for thermal and non-thermal electrons, respectively, and we have put in evidence the dependence on the electrons properties. 
We note that, if we neglect the non-thermal component of the SZE, the SZE will be written as
\begin{equation}
\frac{\Delta I_{tot}(x)}{I_0} = g_{th}(x;T_e^*)y_{th}^* 
\end{equation} 
and, as a result, the determination of the electrons temperature will provide a value $T_e^*$ different from the true one $T_e$.\\
In principle, a different value of the temperature $T_e$ will impact also on the value of the Compton parameter $y_{th}^*$, but we note that 
this parameter is given by the product of the temperature times the optical depth (Colafrancesco et al. 2003), and since 
we are dealing with the results obtained from the stacking of a large number of clusters, therefore the value of the average optical depth, and as a consequence the value of $y_{th}$, is just a free parameter when performing this analysis.

For this reason, we estimate the contribution of the relativistic corrections to the thermal SZE and of the non-thermal SZE from the ratio of the signal at two frequencies $x_1$ and $x_2$. The ratio of the total signal at the two frequencies is given by:
\begin{equation}
\frac{\Delta I(x_1)}{\Delta I(x_2)} = \frac{g_{th}(x_1;T_e)y_{th} + g_{nt}(x_1;s_e,p_1)y_{nt}}{g_{th}(x_2;T_e)y_{th} + g_{nt}(x_2;s_e,p_1)y_{nt}}
\label{ratio1}
\end{equation}
while the value of the same ratio obtained by neglecting the non-thermal contribution is
\begin{equation}
\frac{\Delta I(x_1)}{\Delta I(x_2)} = \frac{g_{th}(x_1;T_e^*)y_{th}^*}{g_{th}(x_2;T_e^*)y_{th}^*} = \frac{g_{th}(x_1;T_e^*)}{g_{th}(x_2;T_e^*)} \, .
\label{ratio2}
\end{equation}
By defining
\begin{equation}
R \equiv \frac{g_{th}(x_1;T_e^*)}{g_{th}(x_2;T_e^*)}
\end{equation}
and
\begin{equation}
X=\frac{y_{nt}}{y_{th}} ,
\end{equation}
and equating eqs.(\ref{ratio1}) and (\ref{ratio2}), we obtain
\begin{equation}
\frac{y_{th}(g_{th}(x_1;T_e) + X g_{nt}(x_1;s_e,p_1))}{y_{th}(g_{th}(x_2;T_e) + X g_{nt}(x_2;s_e,p_1))} = R \, ,
\end{equation}
from which we obtain
\begin{equation}
X = \frac{R g_{th}(x_2;T_e)-g_{th}(x_1;T_e)}{g_{nt}(x_1,s_e,p_1)- R g_{nt}(x_2;s_e,p_1)} \, .
\label{value.x}
\end{equation}

\section{Results}

We apply eq.(\ref{value.x}) to the results presented in H16; we specifically refer to the results presented in Figure 4 of H16, where the temperature obtained from the SZE is compared to the one obtained from X-ray spectroscopic measurements. We consider that this case is more solid w.r.t. the other case presented in Figure 3 of H16, where the temperature obtained from the scaling relation between temperature and X-ray luminosity is used. We also work at the frequencies of 545 and 353 GHz (i.e. $x_1\sim9.59$ and $x_2\sim6.21$), where the non-thermal contribution for high-temperature clusters is expected to be more evident; we note also that H16 observed that the ratio of the SZE signal at these two frequencies is changing with the temperature because of the relativistic corrections; therefore we consider that this ratio is a good indicator of the amount of relativistic corrections. 

We consider the temperature bin where the discrepancy between X-ray and SZE obtained temperatures is more evident, i.e. $k_B T_X=$ 11 keV in Figure 4 of H16, and assume that the cluster real temperature is $k_BT_e=k_BT_X$, while we fix $k_BT_e^*=k_BT_{SZ}$, i.e. 15 keV.

For the non-thermal electrons population, we choose - as a reference case - a population similar to the best case found from the SZE data analysis in the Bullet cluster, i.e. with spectral index $s_e=3.7$ (Marchegiani \& Colafrancesco 2015), and calculate the value of $X$ by changing $p_1$. We will discuss later the effect of changing the value of the spectral index. 

\begin{figure}
\begin{center}
{
 \epsfig{file=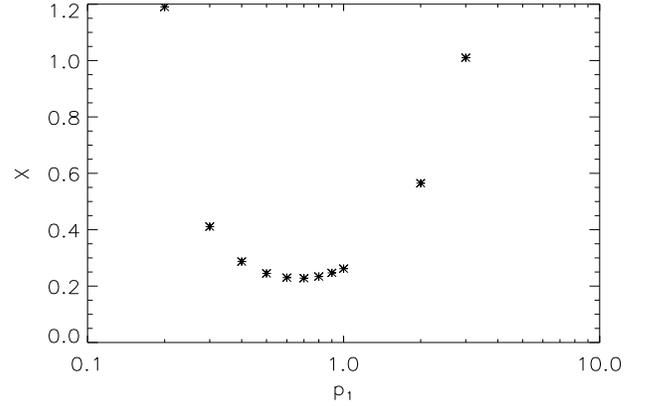,height=6.0cm,width=9.0cm,angle=0.0}
}
\end{center}
 \caption{Values of the pressure ratio $X=y_{nt}/y_{th}$ required to explain the discrepancy between the SZE and X-ray derived cluster temperatures in term of a non-thermal SZE component, calculated according to eq.(\ref{value.x}) for a non-thermal population with $s_e=3.7$ and minimum momentum $p_1$, and working using the ratio of the SZE signals at the frequencies of 545 and 353 GHz.}
 \label{fig.Xvsp1}
\end{figure}

We found that the condition $X<1$ (i.e. the non-thermal pressure is lower than the thermal one) is satisfied only in a narrow range of values of $p_1$, i.e. $0.2<p_1<3$ (see Fig.\ref{fig.Xvsp1}). In this range the minimum possible value of $X$ is obtained for $p_1=0.7$, where $X=0.228$. Including the error bars on $T_{SZ}$ shown in the plot of H16 we obtain the value $X=0.228\pm0.012$ for $p_1=0.7$.

We must stress that the uncertainties on the properties of the non-thermal electron population produce an uncertainty on the strength of these estimates. By changing the value of $s_e$, the value of $X$ that can be derived is changed, as well as the corresponding value of $p_1$, even if this last quantity changes by a small factor: for example, for $s_e=2.7$ we find that the minimum possible value of $X$ is obtained for $p_1=0.5$, for which $X=0.341\pm0.018$.
Anyway we note that this value is of the same order of magnitude w.r.t. the case with $s_e=3.7$, suggesting that, if the non-thermal SZE interpretation of these data is correct, a pressure ratio of the order of $20-30\%$ is a quite strong requirement. 

We stress also that these results are obtained by working only with the ratio of the signal at two frequencies (i.e. 353 and 545 GHz),
and using only the final results of the analysis of H16, where the discrepancy between the X-rays and SZE determined temperatures was suggested only at $\sim1.5\sigma$ level; once this discrepancy will be detected at a higher statistical significance, in order to obtain more precise results on the possibility that the data indicate a contribution from the non-thermal SZE, and on the determination of the parameters of the non-thermal electrons in galaxy clusters (i.e. $X$ as well as $s_e$ and $p_1$ separately) it would be necessary to reanalyze the whole sample of clusters by considering the contribution of the non-thermal SZE at all the frequencies considered. This work is outside the scope of this paper and will be considered in a future work. 

An even better possibility would be to measure the SZE signal at a high number of frequencies along the whole range of frequencies until $\sim 1$ THz, because in this way it would be possible to derive much more precise constraints on both the thermal and non-thermal SZE (Colafrancesco \& Marchegiani 2010); an instrument like Millimetron (Kardashev et al. 2014) will allow this analysis.

\section{Discussion and conclusions}

We have explored the possibility that the discrepancy between the SZE and X-ray estimated values of the cluster temperature suggested by H16 is due to the presence of a physical non-thermal SZE component instead of calibration issues. The main result of our analysis is that the non-thermal SZE can explain this discrepancy under the requirement that the non-thermal electrons have a very low value of the minimum momentum, of the order of $p_1=0.5-1$. This result is in agreement with the results obtained from the SZE analysis of the Bullet cluster (Colafrancesco et al. 2011, Marchegiani \& Colafrancesco 2015); however, this condition is hard to be maintained in a large number of clusters, considering that electrons with this low energy are expected to have a very small lifetime because of energy losses due to non-thermal bremsstrahlung and Coulomb interactions with the thermal gas (see Sarazin 1999), and also considering that the cluster sample here considered should contain mainly clusters without radio halo (or with low power radio halos), where the amount of non-thermal electrons is expected to be lower. 

In addition, the derived ratio between non-thermal and thermal pressures, of the order of $20-30\%$, is an issue for our knowledge of the physics of non-thermal processes in galaxy clusters. In fact, from the upper limits derived from a stacked analysis of Fermi-LAT data in galaxy clusters, Huber et al. (2013) derived an upper limit of the order of $5\%$ for the ratio between non-thermal protons and thermal energy contents. Since processes accelerating cosmic rays in galaxy clusters are believed to provide an energy ratio of the order of 0.01 or smaller between electrons and protons, there would be the problem to find a source of electrons acceleration that is selectively not effective for protons, that is an analogous problem to the one derived from the comparison between radio halos properties and the Fermi-LAT upper limits in galaxy clusters (Vazza et al. 2016).
However, we note that from a statistical study of SZE and X-ray properties of a sample of 23 radio halo clusters an average value of $X\sim0.55$ was found (Colafrancesco et al. 2014), and that in the Bullet cluster the SZE data seem to indicate values of $p_1=1$ and $X=91\%$ (Marchegiani \& Colafrancesco 2015). Therefore these problems seem to be present in all the cases where the SZE is used as a probe of the non-thermal electrons in galaxy clusters, suggesting that something might be still missing in our understanding of the properties of non-thermal phenomena in galaxy clusters, or alternatively that this way to use the SZE measured at a very limited number of frequencies, when instead it would be necessary to use a high number of frequencies (Colafrancesco \& Marchegiani 2010), is not sufficiently good to obtain reliable information on the non-thermal content in galaxy clusters.

We note here that the solution recently suggested for radio halos, i.e. electrons produced by Dark Matter annihilation (Marchegiani \& Colafrancesco 2016), is not working effectively for the SZE signal, because the DM-produced electrons do not extend their spectrum at low energies following a simple power-law (their spectrum is quite flatter), and hence do not produce a significant SZE signal (Marchegiani \& Colafrancesco 2015).

A possible solution is that the electrons producing the non-thermal SZE are not completely diffused into the ICM, but are mainly confined in bubbles of relativistic plasma where the density of thermal gas is low, as observed in several clusters (see, e.g., Fabian et al. 2006). In this case the small density of the target thermal nuclei would allow the low-energy non-thermal electrons to survive for a longer time and to maintain a power-law spectral shape down to low values of $p_1$. 
The injection of electrons from radio galaxies jets and the formation of lobes seems to be the most likely mechanism to produce this kind of structures. If the jets are mainly leptonic, or if some mechanism can maintain the protons confined in these bubbles, we also do not expect a strong gamma-ray emission to be produced in the bubbles because of the lack of target nuclei.
However, if these bubbles have an internal pressure similar to the pressure of the surrounding ICM, to obtain a value of $X\sim20-30\%$ it would be necessary that the bubbles cover a volume of the order of a quarter of the cluster volume, a condition that might be difficult to obtain. This problem can be partially alleviated if a strong contribution comes from bubbles of relativistic plasma located in the outskirts of the cluster, like giant radiogalaxies lobes or radio relics.

We conclude that, once a higher significance detection of the discrepancy between X-rays and SZE estimated temperatures will be obtained, there are two ways to derive stronger conclusions from this analysis: the first one is to reproduce the full analysis done in H16 by considering the additional contribution of the non-thermal SZE over a wide range of frequencies and considering a number of possible models for the non-thermal electrons, with the aim to determine not only the pressure but also the spectral properties of the non-thermal electrons. 
Another possibility is to perform multi-frequency and spatially-resolved observations of the SZE in galaxy clusters, and to compare the results with X-ray and radio observations, in order to check if the non-thermal SZE is mainly produced in bubbles with high non-thermal and low thermal plasma content. Instruments like Millimetron (see, e.g., Kardashev et al. 2014) are suitable to this goal. 

Another issue that will need to be considered is the effect of temperature non-uniformities within the clusters, that can create a bias in the estimation of the mean temperature of a cluster when considered as isothermal. For example, it has been shown that the X-ray temperature determinations based on a single temperature fit are biased towards high-density and low-temperature regions (e.g. Mazzotta et al. 2004), because the emissivity of the thermal bremsstrahlung is proportional to $n_e^2 \sqrt{T_e}$. The thermal SZE is, in the non-relativistic formulation, linearly proportional to both $n_e$ and $T_e$, and therefore one should expect than the SZE derived temperatures are less biased than the X-ray derived ones. However, the relativistic formulation makes the dependence on $T_e$ non-linear, and this translates in a bias towards high values of the temperature. As in the case of the X-ray analysis, a spatially resolved measure of the temperature along the cluster through the SZE can allow to recover the true distribution of the temperature through a de-projection technique (Colafrancesco \& Marchegiani 2010). However, when working with the spatially integrated SZE values in clusters that are assumed to be isothermal, the bias introduced by a single temperature fit to the SZE can be high, as well the variance on the temperature, that, expecially in disturbed clusters where a big range of temperatures is present, can have values of the order of the average temperature (Prokhorov \& Colafrancesco 2012).

As an example in a simple and idealized case, Vikhlinin (2006) found that, assuming a cluster with two thermal populations with the same bremsstrahlung emission measures and temperatures of 4 and 16 keV, the resulting X-ray emission can be fitted by a single thermal population with temperature of $\sim7$ keV. In order to compare the SZE bias with this result, we obtained an analogous estimate of the temperature derived from a single temperature fit to the SZE produced by two populations with the same optical depth and temperatures of 4 and 16 keV, and we found that the SZE derived temperature is $\sim13$ keV. Therefore, in this case both these estimates deviate from the average temperature of 10 keV by a factor of the order of $\sim30\%$ in opposite directions, that would be sufficient to account for all the claimed discrepancy as resulting from H16. However, when working with the stacked analysis of a large sample of clusters that can have different structures of the temperatures (including e.g. temperature gradients, cooling flows, shock waves, cold fronts), the estimate of the resulting bias on the discrepancy between X-ray and SZE derived temperatures is a complex issue and will be addressed in a future paper.

\section*{Acknowledgments}

We thank the Referee for several useful comments and suggestions.
This work is based on the research supported by the South African Research Chairs Initiative of the Department of Science and Technology and National Research Foundation of South Africa (Grant No 77948).
P.M. acknowledges support from the DST/NRF SKA post-graduate bursary initiative.

\bsp

\label{lastpage}

\end{document}